# High-Spatial-Resolution Optical Correlation-Domain Reflectometry with 100-km Measurement Range

Takaki Kiyozumi, Soshi Yoshida, Yuta Higa, Keisuke Motoda,
Sze Yun Set, Shinji Yamashita, and Yosuke Mizuno

## I. INTRODUCTION

Optical fiber communication underpins modern information infrastructure. As network traffic continues to increase, efficient maintenance and management of optical fiber links have become more critical, and optical reflectometry has become an indispensable diagnostic technique [1][2]. In this context, a central challenge in optical reflectometry is to realize low-cost, high-speed measurement systems that can profile optical loss and locate fault (damage) points along a fiber under test (FUT) over distances on the order of ~100 km, corresponding to a typical repeater span. Existing optical reflectometry techniques are broadly classified into optical time-domain reflectometry (OTDR), optical frequency-domain reflectometry (OFDR), and optical correlation-domain reflectometry (OCDR).

Among these techniques, OTDR is the most widely deployed in practical field applications. In OTDR, optical pulses are launched into the FUT, and the positions of reflective events are obtained from the round-trip propagation time of the backscattered or reflected light [3][4]. Although OTDR offers a relatively simple system configuration and supports long measurement ranges, improving spatial resolution requires shortening the pulse width, which reduces the injected optical energy and degrades the signal-to-noise ratio (SNR). To compensate for this trade-off, extensive temporal averaging is required. Consequently, for long-range measurements beyond ~100 km, the acquisition time can extend to several minutes, while the attainable spatial resolution is typically limited to the meter level [6]. To alleviate these limitations, high-resolution OTDR schemes based on photon counting [6][8] and optical sampling [7] have been proposed. However, these approaches still suffer from constraints such as measurement distances restricted to a few kilometers or difficulty in accurately observing optical loss originating from Rayleigh backscattering.

In OFDR, the optical frequency of the probe light is linearly swept, and the positions of reflection points are identified from the beat frequency between the light reflected from the FUT and a reference beam[9][10]. OFDR offers very high spatial resolution, ranging from several micrometers to several centimeters[11]. However, it is difficult to perform measurements beyond the coherence length of the light source, which fundamentally limits the measurable distance [12]. To overcome this limitation, phase-noise-compensated OFDR (PNC-OFDR) [13], which digitally compensates laser phase noise, and time-gated OFDR (TGD-OFDR) [14], which extends the measurement range through time gating, have been proposed. These techniques realize meter-scale spatial resolution over measurement ranges of several tens of kilometers. On the other hand, they require sophisticated digital signal processing, auxiliary interferometers, and narrow-linewidth wavelength-swept light sources, which increase system complexity and overall cost.

OCDR is a distributed fiber-optic measurement technique based on the synthesis of an optical coherence function (SOCF) [15]. By applying sinusoidal frequency modulation to continuous-wave light and launching it into the FUT, an optical correlation peak is formed at a specific position along the fiber [16]. Around this peak, the heterodyne detection signal is selectively enhanced, and by scanning the correlation peak along the FUT, distributed measurements over the entire length become possible. A distinctive feature of OCDR is that it can achieve measurement ranges exceeding the coherence length of the light source [17][18]. Furthermore, because the frequency modulation can be implemented by directly modulating the light source, the system configuration can be kept relatively simple. In addition, OCDR provides random access to arbitrary positions along the FUT, enabling local high-speed measurements[19]. Owing to these advantages, OCDR is regarded as a promising approach to simultaneously addressing the short measurement range and high cost of OFDR, as well as the long acquisition time of OTDR.

Despite these advantages, existing OCDR systems still exhibit a trade-off between measurement range and spatial resolution. Although Rayleigh backscattering in FUTs with lengths of several tens of kilometers has been observed, the spatial resolution has remained on the order of several meters [21][22].

In this study, we aim to mitigate this trade-off in OCDR by optimizing the modulation waveform. We demonstrate profiling of optical loss and fault locations in a ~100-km-long FUT with a spatial resolution of ~19 cm. The principle of the proposed method and the corresponding experimental results are presented in the following sections.

## II. PRINCIPLE

### A. OCDR using a sinusoidal wave

The basic configuration of the OCDR system is shown in Fig. 1. The light reflected from the fiber under test (FUT) and the reference light interfere in a Michelson-interferometer configuration. The incident light is frequency-modulated by a sinusoidal wave, and its complex electric field is expressed as

$$E_s(t) = \exp\left(2\pi \int_0^t f(t')dt'\right), \qquad (1)$$

where the instantaneous optical frequency $f(t')$ is given by

$$f(t') = f_0 + \Delta f \sin(2\pi f_m t'). \quad (2)$$

Here, $f_0$ is the center optical frequency, $\Delta f$ is the frequency-modulation amplitude, and $f_m$ is the modulation frequency.

The beat spectrum $S_A(x, f)$ between the reference light and the light reflected from the position located at a distance $x$ [m] from the zero optical path difference point (ZOPD) is given by

$$S_A(x, f) = \sum_{\nu=-\infty}^{\infty} \left| J_\nu \left( 2\frac{\Delta f}{f_m} \sin\left(\frac{2\pi f_m n x}{c}\right) \right) \delta(f + \nu f_m) \right|. \quad (3)$$

where $J_\nu(\cdot)$ is the $\nu$-th-order Bessel function of the first kind, $\delta(\cdot)$ is the Dirac delta function, $c$ is the speed of light in vacuum, and $n$ is the refractive index of the optical path. When a silica optical fiber is used, $n \approx 1.46$. In Eq. (3), the variation of the spectrum at the frequency

$$\gamma(x) = \left| J_0 \left( 2\frac{\Delta f}{f_m} \sin\left(\frac{2\pi f_m n x}{c}\right) \right) \right|. \quad (4)$$

This function $\gamma(x)$ is referred to as the synthesized optical coherence function (SOCF), and its shape coincides with that of the reflection peak. Equation (4) shows that $\gamma(x)$ is periodic with respect to the distance $x$ from the ZOPD. In each period, there is one "correlation peak," which corresponds to the maximum of the interference distribution. The interval between neighboring correlation peaks is

$$d_m = \frac{c}{2n f_m}. \quad (5)$$

The reflected light from positions corresponding to correlation peaks is clearly detected, whereas the reflected light from other positions is strongly suppressed. Therefore, each correlation peak acts as a measurement point. This property, which enables selective measurement of arbitrary locations along the FUT, is called "random accessibility."

The distance $x_{\text{cp}}$ of a correlation peak from the ZOPD is given by

$$x_{cp} = K \frac{c}{2n f_m}, \quad (6)$$

where $K$ represents the order of the correlation peak: the zero-order peak is located at the ZOPD, followed by the first-, second-, and higher-order peaks. By adjusting the modulation frequency $f_m$, the correlation peaks can be shifted along the FUT. The position of the zero-order correlation peak is adjusted by inserting a delay line into one arm of the interferometer. When multiple correlation peaks exist within the FUT, it becomes difficult to identify the measurement point. Therefore, the measurement range of OCDR is limited to the interval between adjacent correlation peaks, which is given by Eq. (6).

The spatial resolution $\Delta z$ is defined as the full width at half maximum (FWHM) of a correlation peak and is

$$\Delta z \cong \frac{0.76 c}{\pi \Delta f n}. \quad (7)$$

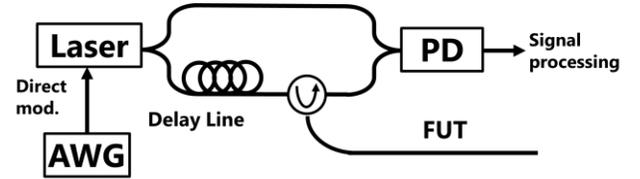

Fig. 1 Conceptual setups of the OCDR.

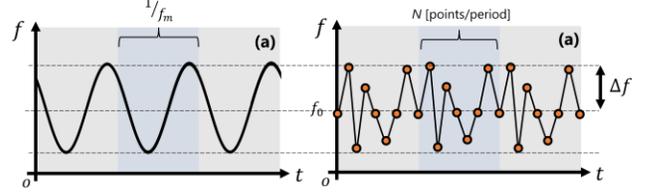

Fig. 2 Modulation waveforms: (a) sinusoidal, (b) PPRM.

### B. OCDR using Gaussian noise

When the sinusoidal frequency modulation is replaced with white zero-mean white Gaussian noise process $G(t)$, the complex electric field of the incident light is expressed as

$$E_n(t) = \exp\left( 2\pi \int_0^t G(t') dt' \right). \quad (8)$$

We now consider the beat spectrum obtained from the interference between the reference light and the delayed light. Let $\tau_d = \frac{2nx}{c}$ denote the time delay corresponding to the optical path difference between the two arms of the interferometer. The beat spectrum $S_d(f, \tau_d)$ as a function of the electrical frequency $f$ and the delay time $\tau_d$ is given by [23]

$$\begin{aligned} S_d(f, \tau_d) &= e^{-2\pi \delta f \tau_d} \delta(f) \\ &+ \frac{\delta f}{\pi(f^2 + \Delta f^2)} \\ &\left\{ 1 - e^{-2\pi \delta f \tau_d} \left( \cos(2\pi f \tau_d) - \frac{f}{2\Delta f} \sin(2\pi f \tau_d) \right) \right\} \\ &- \frac{1}{2\pi^2 f} e^{-2\pi \delta f \tau_d} \sin(2\pi f \tau_d). \end{aligned} \quad (9)$$

where $\Delta f$ is the half-width of the Gaussian noise spectrum, $\delta(\cdot)$ is the Dirac delta function, and $f$ is the electrical frequency. From Eq. (9), the variation of the spectrum at $f = 0$ can be written as

$$|\gamma(\tau_d)| = \frac{1}{2\pi \Delta f} + \left( 1 - \frac{1}{2\pi \Delta f} - \frac{\tau_d}{\pi} \right) e^{-4\pi \Delta f \tau_d} \quad (10)$$

This function $\gamma(\tau_d)$ represents the synthesized optical coherence function (SOCF) obtained when white Gaussian noise modulation is used. The FWHM of the corresponding correlation peak in the spatial domain is

$$\Delta z \cong \frac{c \ln 2}{4\pi n \Delta f} \quad (11)$$

Comparing Eq. (11) with the sinusoidal modulation case (Eq. (7)), it is found that, for the same modulation amplitude $\Delta f$, the linewidth obtained with Gaussian-noise modulation can be significantly narrower. In practical implementations, the Gaussian noise is digitized and converted into a periodic waveform with period $1/f_m$, as shown in Fig. 2 (b), where $f_m$ is the repetition frequency of the digitized noise pattern.

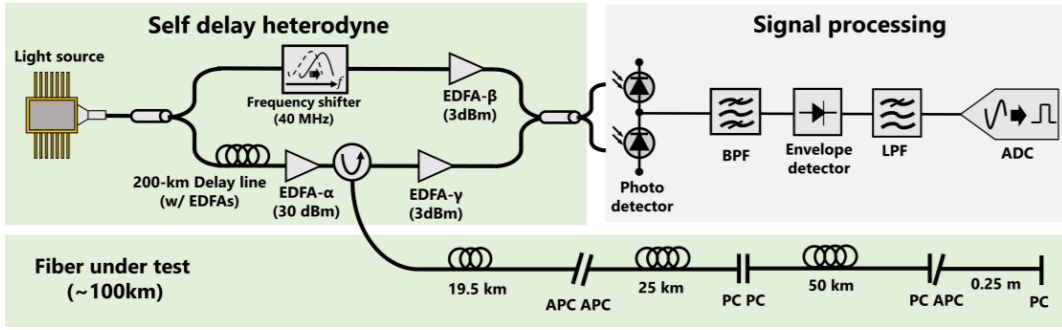

Fig. 3 Experimental setup of the OCDR system. EDFA: erbium-doped fiber amplifier; BPF: band-pass filter; LPF: low-pass filter; ADC: analog-to-digital converter; APC: angled physical-contact connector; PC: physical-contact connector.

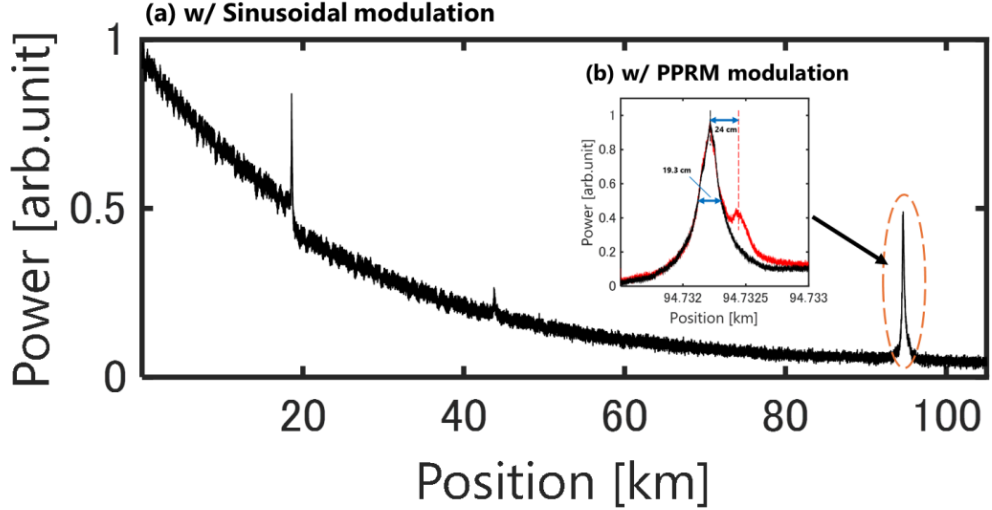

Fig. 4 Experimental results under two modulation schemes: (a) full-length measurement of the FUT using sinusoidal modulation; (b) end-reflection measurement of the FUT using PPRM modulation.

### III. EXPERIMENTS

An experimental demonstration of spatial-resolution enhancement using periodic pseudo-random modulation (PPRM) was carried out with the optical correlation-domain reflectometry (OCDR) system shown in Fig. 3. The light source was a butterfly-type laser diode operating at a center wavelength of 1550 nm with a linewidth of 20 MHz. The emitted light was split by an optical coupler into a reference arm and a probe arm coupled into the FUT. A 200-km delay line was inserted in the probe arm to adjust the position of the ZOPD. To compensate for the insertion loss, four erbium-doped fiber amplifiers (EDFAs) were employed, and the final stage, denoted EDFA-$\alpha$, increased the optical power up to 30 dBm. In the reference arm, a 40-MHz acousto-optic modulator (AOM) was inserted for heterodyne detection.

The light reflected from the FUT was amplified to 3 dBm by EDFA-$\gamma$, and the reference light was amplified to 3 dBm by another amplifier (EDFA-$\beta$). These two beams were then mixed in a balanced photodiode (BPD) for heterodyne detection. The electrical signal from the BPD was first passed through a band-pass filter with a center frequency of 40 MHz and a bandwidth of 3 MHz to extract the beat component. It was then passed through a 1-Hz low-pass filter for noise suppression and finally digitized by an analog-to-digital converter (ADC) operating at 1.25 kS/s.

The FUT was a ~100-km single-mode fiber (SMF). Reflection points were introduced at ~20 km, ~45 km, and ~95 km from the circulator output (defined as 0 m) by using an angled physical contact/angled physical contact (APC/APC) connector, a physical contact/physical contact (PC/PC) connector, and a PC open end, respectively.

First, to rapidly identify the locations of all reflection points along the entire length of the FUT, sinusoidal frequency modulation was applied to the incident light as described in Eq. (2). By sweeping the modulation frequency $f_m$ from 1050 to 300 Hz, the fist correlation peak was scanned over the whole fiber within 10 s, according to Eq. (6).

Next, to perform random-access measurements only near the fiber end, PPRM was applied as illustrated in Fig. X. The modulation parameters were set to a random-modulation amplitude $\Delta f \simeq 5$ Hz and $N = 5000$ discrete points per period. By sweeping the modulation frequency $f_m$ from 520.2725 to 520.2770 Hz, the correlation peak was scanned over a limited region of ~1.5 m around the FUT end. To evaluate the spatial-resolution performance, random-access measurements were also carried out under the same conditions with a 25-cm patch cable connected to the FUT end. Both sinusoidal and PPRM modulations were implemented by directly modulating the drive current of the laser source.

## IV. Results and discussion

Figure 4(a) shows the measurement results over the entire length of the FUT, where distinct reflection peaks can be observed at the positions corresponding to reflection points A, B, and C. In addition, optical attenuation due to Rayleigh scattering is observed throughout the entire trace.

The red curve in Fig. 4(b) presents the result of the random-access measurement performed near the end of the FUT. From the width of the reflection peak at the fiber end, it is evident that the position can be identified with an accuracy of ~19 cm. The black curve in Fig. 4(b) shows the result obtained when a 25-cm optical fiber was connected to the end of the FUT. Two separated peaks are clearly observed, demonstrating a spatial resolution of at least 25 cm.

These experimental results confirm that the proposed method successfully detected the reflection points (connection points) along a ~100-km FUT with an accuracy of 19 cm in a total measurement time of 20 seconds.

## V. Conclusion

In this paper, we propose an OCDR employing PPRM modulation and demonstrate that it can acquire the reflection points and optical loss distribution in a ~100-km FUT with a spatial resolution of about 25 cm and a measurement time of about 10 s. The system consists only of a low-cost DFB laser, a single interferometer, and analog signal-processing circuitry, and can therefore be implemented with a simple and low-cost configuration. These results show that the proposed PPRM-OCDR is a promising technique that can significantly improve maintenance efficiency in future optical fiber networks.

## Reference

[1] Agrawal, G. P. *Fiber-Optic Communication Systems*, 5th edn (Wiley, 2021).
[2] Bai, Q., Feng, Z. & Zhang, L. Recent advances in Brillouin optical time domain reflectometry. *Sensors* **19**, 1862 (2019).
[3] Barnoski, M. K., Rourke, M. D., Jensen, S. M. & Melville, R. T. Optical time domain reflectometer. *Appl. Opt.* **16**, 2375–2379 (1977).
[4] Personick, S. D. Photon probe—an optical-fibre time-domain reflectometer. *Bell Syst. Tech. J.* **56**, 355–366 (1977).
[5] Zhao, Q. *et al.* Long-haul and high-resolution optical time domain reflectometry using superconducting nanowire single-photon detectors. *Sci. Rep.* **5**, 10441 (2015).
[6] Wang, Z. N. *et al.* Long-range and high-precision correlation optical time-domain reflectometry utilizing an all-fiber chaotic source. *Opt. Express* **23**, 15514–15520 (2015).
[7] Eraerds, P. *et al.* Photon counting OTDR: advantages and limitations. *J. Lightwave Technol.* **28**, 952–964 (2010).
[8] Wang, S., Fan, X. & He, Z. Ultrahigh resolution optical reflectometry based on linear optical sampling technique with digital dispersion compensation. *IEEE Photon. J.* **9**, 6804710 (2017).
[9] MacDonald, R. I. Frequency domain optical reflectometer. *Appl. Opt.* **20**, 1840–1844 (1981).
[10] Eickhoff, W. & Ulrich, R. Optical frequency domain reflectometry in single-mode fibre. *Appl. Phys. Lett.* **39**, 693–695 (1981).
[11] Ghafoori-Shiraz, H. & Okoshi, T. Fault location in optical fibres using optical frequency domain reflectometry. *J. Lightwave Technol.* **4**, 316–322 (1986).
[12] Ding, Z. *et al.* Distributed optical fiber sensors based on optical frequency domain reflectometry: a review. *Sensors* **18**, 1072 (2018).
[13] Fan, X., Koshikiya, Y., Ito, F., He, Z. & Hotate, K. Phase-noise-compensated optical frequency domain reflectometry with measurement range beyond laser coherence length realized using concatenative reference method. *Opt. Lett.* **32**, 3227–3229 (2007).
[14] Liu, Q., Fan, X. & He, Z. Time-gated digital optical frequency-domain reflectometry with 1.6 m spatial resolution over entire 110 km range. *Opt. Express* **23**, 25988–25997 (2015).
[15] Hotate, K. & Kamatani, O. Optical coherence-domain reflectometry by synthesis of coherence function. *J. Lightwave Technol.* **11**, 1701–1710 (1993).
[16] Hotate, K., Okugawa, T. & Saida, T. Applications of synthesis of the optical coherence function: one-dimensional reflectometry and three-dimensional optical image processing. *Proc. SPIE* **2576**, 66–77 (1995).
[17] Kashiwagi, M. & Hotate, K. Long range and high resolution reflectometry by synthesis of optical coherence function at region beyond the coherence length. *IEICE Electron. Express* **6**, 497–503 (2009).
[18] Kiyozumi, T., Noda, K., Zhu, G., Nakamura, K., Set, S. Y. & Mizuno, Y. Beat spectrum in dual-laser optical correlation-domain reflectometry. Proc. SPIE 13639, 136393N (2025).
[19] Hotate, K. & He, Z. Synthesis of optical coherence function and its applications in distributed and multiplexed optical sensing. *J. Lightwave Technol.* **24**, 2541–2557 (2006).
[20] Kiyozumi, T., Noda, K., Zhu, G., Nakamura, K. & Mizuno, Y. Modified expression for spatial resolution in optical correlation-domain reflectometry. *IEEE Trans. Instrum. Meas.* **73**, 1–11 (2024).
[21] Miyamae, T. *et al.* Observation of Rayleigh scattering by simplified optical correlation-domain reflectometry without frequency shifter. *Appl. Phys. Express* **16**, 052004 (2023).
[22] Xue, Y., Niu, Y. & Gong, S. External modulation optical coherent domain reflectometry with long measurement range. *Sensors* **21**, 5510 (2021).
[23] Okoshi, T. & Kikuchi, K. *Coherent Optical Fiber Communications*. Springer, Dordrecht (1988).

**Takaki Kiyozumi** (Student Member, IEEE) received the B.E. degree in 2020 and the M.E. degree in electrical and computer engineering in 2024 from Yokohama National University, Yokohama, Japan.

Since 2024, he has been studying distributed fiber-optic sensing for his Dr. Eng. degree at the University of Tokyo, Tokyo, Japan.

Mr. Kiyozumi is the winner of the IEEE IM Japan Chapter Student Award 2021, the Outstanding Student Award 2021 from the Optical Fiber Technologies Conference (OFT), the Institute of Electronics, Information and Communication Engineers (IEICE) of Japan, and the Best Student Paper Award from the 27th Optoelectronics and Communications Conference (OECC 2022). He is a student member of the IEEE and the Japanese Society of Applied Physics (JSAP).

**Soshi Yoshida** is currently pursuing his M.E. degree in electrical engineering at the University of Tokyo, Tokyo, Japan. His research concentrates on distributed fiber-optic sensing techniques, with a particular focus on optical correlation-domain reflectometry. He was awarded the Student Poster Award at the Optical Fiber Technologies Conference (OFT) in 2024. He is a student member of the Japanese Society of Applied Physics (JSAP).

**Yuta Higa** is currently pursuing his M.E. degree in electrical and computer engineering at Yokohama National University, Yokohama, Japan. His academic focus is on distributed fiber-optic sensing techniques, with a particular interest in optical correlation-domain reflectometry. He is a student member of the Japanese Society of Applied Physics (JSAP).

**Sze Yun Set** (Senior Member, IEEE) received the B.Eng. first-class Honours degree from the University of Southampton, Southampton, U.K., in 1993, and the Ph.D. from the Optoelectronics Research Centre of the same university in 1999.

From 1998 to 2001, he was a JSPS Postdoctoral Fellow with the Research Center for Advanced Science and Technology (RCAST), the University of Tokyo, Tokyo, Japan. He was appointed as the CEO and CTO of the Alnair Labs Corporation in 2005. He has been with the University of Tokyo since 2016 and is currently a Project Professor. His research interests include short-pulsed fiber lasers, carbon-nanotube/graphene photonics, LIDAR, 3D laser imaging, bio-imaging, and optical sensing.

Prof. Set has contributed to more than 200 refereed journal and conference publications and 12 patents. He is a Fellow of OPTICA and was a Visiting Fellow at Clare Hall and the Cambridge Graphene Centre, Cambridge University. He was an IEEE Distinguished Lecturer, and a Deputy Director of the Japan Society of Applied Physics (JSAP).

**Shinji Yamashita** (Member, IEEE) received the B.E., M.E., and Dr.Eng. degrees in electronic engineering from the University of Tokyo, Tokyo, Japan, in 1988, 1990, and 1993, respectively.

He was appointed as a Research Associate in 1991, a Lecturer in 1994, an Associate Professor in 1998, and a Professor in 2009, with the University of Tokyo. He is currently a Professor with the Research Center for Advanced Science and Technology (RCAST), the University of Tokyo. From 1996 to 1998, he was with the Optoelectronics Research Center, University of Southampton, Southampton, U.K., as a visiting research fellow. He has been engaged in research of coherent optical fiber communications, optical fiber amplifiers, fiber nonlinearities, and fiber lasers.

Prof. Yamashita has authored/coauthored and presented more than 300 refereed papers in the field. His current research interests include fiber lasers and nonlinear devices for optical fiber communications and sensors. He is a Fellow of OPTICA, and a member of the Institute of Electronics, Information, and Communication Engineers (IEICE) of Japan and the Japan Society of Applied Physics (JSAP).

**Yosuke Mizuno** (Senior Member, IEEE) received the B.E., M.E., and Dr.Eng. degrees in electronic engineering from the University of Tokyo, Tokyo, Japan, in 2005, 2007, and 2010, respectively.

From 2010 to 2012, as a Research Fellow (PD) of the Japan Society for the Promotion of Science (JSPS), he worked on polymer optics at Tokyo Institute of Technology, Japan. In 2011, he stayed at BAM Federal Institute for Materials Research and Testing, Germany, as a Visiting Research Associate. From 2012 to 2020, he was an Assistant Professor at Tokyo Institute of Technology. Since 2020, he has been an Associate Professor at the Faculty of Engineering, Yokohama National University, Japan, where he is active in fiber-optic sensing and polymer optics.

Prof. Mizuno is the winner of the Funai Information Technology Award 2017, the Optics Design Award 2018, and the Young Scientist's Award 2021, Commendation for Science and Technology, Minister of Education, Culture, Sports, Science and Technology (MEXT). He is a member of the IEEE Photonics Society (Senior Member), the Japanese Society of Applied Physics (JSAP), the Optical Society of Japan (OSJ), and the Institute of Electronics, Information, and Communication Engineers (IEICE) of Japan.